# Efficient Compilation and Mapping of Fixed Function Combinational Logic onto Digital Signal Processors Targeting Neural Network Inference and Utilizing High-level Synthesis

SOHEIL NAZAR SHAHSAVANI*, ARASH FAYYAZI*, MAHDI NAZEMI, and MASSOUD PEDRAM,
Ming Hsieh Department of Electrical and Computer Engineering, University of Southern California

Recent efforts for improving the performance of neural network (NN) accelerators that meet today's application requirements have given rise to a new trend of logic-based NN inference relying on fixed function combinational logic. Mapping such large Boolean functions with many input variables and product terms to digital signal processors (DSPs) on Field-programmable gate arrays (FPGAs) needs a novel framework considering the structure and the reconfigurability of DSP blocks during this process. The proposed methodology in this paper maps the fixed function combinational logic blocks to a set of Boolean functions where Boolean operations corresponding to each function are mapped to DSP devices rather than look-up tables (LUTs) on the FPGAs to take advantage of the high performance, low latency, and parallelism of DSP blocks. This paper also presents an innovative design and optimization methodology for compilation and mapping of NNs, utilizing fixed function combinational logic to DSPs on FPGAs employing high-level synthesis flow. Our experimental evaluations across several datasets and selected NNs demonstrate the comparable performance of our framework in terms of the inference latency and output accuracy compared to prior art FPGA-based NN accelerators employing DSPs.

## 1 INTRODUCTION

The embedded digital signal processing (DSP) blocks in modern Field Programmable Gate Arrays (FPGAs) support fast and efficient implementation of a wide variety of logic operations. They support simple Boolean operations as well as complicated arithmetic operations such as multiplication in a single instruction, multiple data (SIMD) scheme. DSPs have evolved to support a wide range of applications requiring significant amounts of Boolean operations that may not even necessarily fit on the available lookup tables (LUTs) on an FPGA. In addition to the vast computation capabilities, DSP blocks support dynamic runtime programmability, which allows a single DSP block to be used as a different computational block in each clock cycle. Vendor synthesis tools provide capabilities to utilize the available resources on FPGAs; however, existing tool flows such as high-level synthesis tools fail to fully exploit the existing capabilities, especially the dynamic programmability of DSPs.

Bajaj et al. [10–14] explore how DSP blocks can be deployed to produce high-throughput computational kernels and how their dynamic programmability can be exploited to create efficient implementations of arithmetic expressions. However, their solution suffers from inefficient mapping when it comes to implementing combinational Boolean functions using DSP blocks. In particular, high-level synthesis (HLS) tools do not support time-shared mapping of operations on available resources and usually rely on backend synthesis tools to efficiently map and schedule the operations on the target architecture.

New applications have arisen that produce large sparse Boolean functions with many input variables and product terms. Examples of such applications are found in [6, 8, 18] where the problem of efficient processing of neural networks is formulated as a Boolean logic minimization problem where ultimately, logic expressions compute output of various filters/neurons. In fact, [8] optimizes a target DNN for a given dataset and maps essential parts of the computation in the DNN to ultra-low-latency, low-cost, fixed-function, combinational logic blocks, which can be implemented using LUTs.

However, because neurons designed for state-of-the-art neural networks include tens to hundreds of inputs, the generated Boolean logic expression is huge. Consider the eighth convolutional layer of the VGG16 neural network [15] trained on the CIFAR-10 dataset [4]. This layer consists of 512 $3 \times 3$ filters that are applied to an input volume of $4 \times 4 \times 256$. Therefore, the number of inputs to each filter is $3 \times 3 \times 256 = 2,304$ while the number of input patches is $4 \times 4 = 16$. Therefore, to realize this layer using the approach presented in [8], hundreds of thousands LUTs are needed. Our experiments show that generated Boolean logic expressions for state-of-the-art networks such as VGG16 cannot be fitted into one FPGA if only LUTs are utilized to process logic expressions.

In this paper, we propose a novel framework to map fixed function combinational blocks to DSPs on the FPGAs. The proposed methodology starts by transforming a neural network specification to a set of optimized fixed function combinational logic (FFCL) blocks using the NullaNet framework [6, 8]. Next, we map each FFCL block to a set of Boolean operations, supported by the DSPs. The Boolean operations are then scheduled to be executed on DPSs and the compiler orchestrates the data movement from/to a host device to/from the FPGA to enable loading/storing input/outputs of each DSP block in each computational cycle. During processing, the input values are transferred to the FPGA, stored in the BRAMs, and loaded to the regsiters associated with each DSP block. DSP blocks then carry out the Boolean operations in parallel and store the output values in registers. Output values for each FFCL block are transferred to pre-determined BRAM blocks and later loaded to the DRAM modules interfacing the FPGA. Utilizing the proposed methodology, the operations associated with any FFCL module, irrespective of the number of Boolean operations and the number of inputs/outputs, can be mapped to and executed by the DSP blocks on an FPGA. Hence, the shortcomings associated with mapping a FFCL module to LUTs on FPGAs due to resource limitations are completely addressed.

## 2 PRIOR WORK

Early papers on the binary neural networks (BNNs), such as XnorNet [9] and BinaryNet [2], advocated network training with binary weights and activations to speedup the inference phase. FINN [19] uses FPGAs for accelerating BNNs and proposes a parameterizable dataflow architecture and optimizations to enable real-time response in embedding applications. Other studies, such as Wang et al. [20], employ FPGA LookUp Table (LUT) blocks as inference operators for BNNs. Their results show that the construction of K-LUT-based networks is robust to high levels of pruning with little or no loss of accuracy, allowing them to achieve much higher area and energy efficiency than the state-of-the-art BNN implementations and various common network models. The main idea is replacing XNOR gate with a K-LUT (i.e., logic expansion) by regrowing some connections that are removed in the pruning phase to compensate for the accuracy. But, they still keep some expensive blocks such as adders.

In a parallel track, the idea of converting certain layers of DNNs to fixed-function, combinational logic blocks followed by the mapping of those blocks to look-up tables (LUTs) has been previously discussed in NullaNet [8] and LogicNets [18]. They convert a quantized neural network into hardware building components (i.e., LUTs) that can be efficiently mapped to hardware. NullaNet [8] presents a approach by enumerating all possible input combinations, recording encountered output values, and optimizing truth tables reflecting those input-output combinations of each filter/neuron. LogicNets [18] is able to construct scalable architectures that allow for very high clock frequencies with a competitive accuracy by describing neurons in a network as truth tables by limiting the count of quantized inputs/outputs for each neuron. After that, the authors of NullaNet proposed NullaNet Tiny [6] to outperform the LogicNets in similar settings on the same benchmarks. NullaNet Tiny [6] stands apart from previous work by using various activation functions for different layers to achieve greater performance.

In most cases including those used in this study, the accuracy drop for binary implementation is less than 4%. However, these approaches cannot fit a large neural network like VGG16 on a single FPGA. Because neurons designed for state-of-the-art neural networks include tens to hundreds of inputs, the generated Boolean logic expression is huge and cannot be fitted into one FPGA if only LUTs are utilized to process logic expressions. So, this paper aims to tackle the shortcomings associated with mapping a FFCL module to LUTs on FPGAs due to resource limitations. Evidently, by leveraging a hybrid implementation, i.e., mapping some FFCL modules to LUTs and others to DSPs, a high-performance inference engine for any network on any FPGA device can be achieved.

## 3 PRELIMINARIES AND NOTATIONS

Hereafter, we first describe the general FPGA layout and the principles of SDAccel Environment. We then define the terminology and notations used in the rest of this paper.

### 3.1 FPGA Layout and SDAccel Environment

In this section, we briefly review the overall structure of memory/computational logic on FPGAs and provide an introduction to SDAccel development environment. In a typical FPGA, primitives (low-level blocks), including digital signal processing (DSP) modules, block RAMs (BRAM, on-chip memory components widely used in FPGAs), recently-added Ultra RAMs (URAM, dual-port, single clocked, cascadable synchronous memory blocks available in UltraScale+ FPGAs), and configurable logic blocks (CLB, consists of the Multiple lookup tables, LUTs) are laid out in a columnar fashion. As depicted in Fig. 1, a DSP column (comprised of multiple DSP blcoks), a CLB column, a BRAM column, and another CLB column are laid out in a columnar fashion.

SDAccel is an OpenCL application development environment for FPGA-based Xilinx accelerator cards. The SDAccel environment provides a framework for developing and implementing accelerated FPGA applications using standard high-level programming languages. In the SDAccel framework, an application is divided between a host application and a hardware-accelerated kernel with a high-speed communication channel between them. A host application, which is typically written in C/C++, runs on the host device (i.e., the CPU) with support from API abstractions (such as OpenCL), while the accelerated kernel, written in either C/C++ or Verilog, runs on the FPGA. The communication between the host CPU and the FPGA accelerator board is carried out through the PCIe bus. The host code also provides an interface to enable data transfer to/from the host from/to the kernel conforming with the OpenCL programming paradigm. The host code is subdivided into three main parts–functions that (a) configure the development environment and communication channels, (b) allocate computational resources and construct computational kernels to carry out the actual computations, and (c) post-process the computed values and mange the allocated hardware resources. The orchestration among different functions implements the full stack from high-level description of an application to low-level hardware level implementation using computational resources on an FPGA fabric.

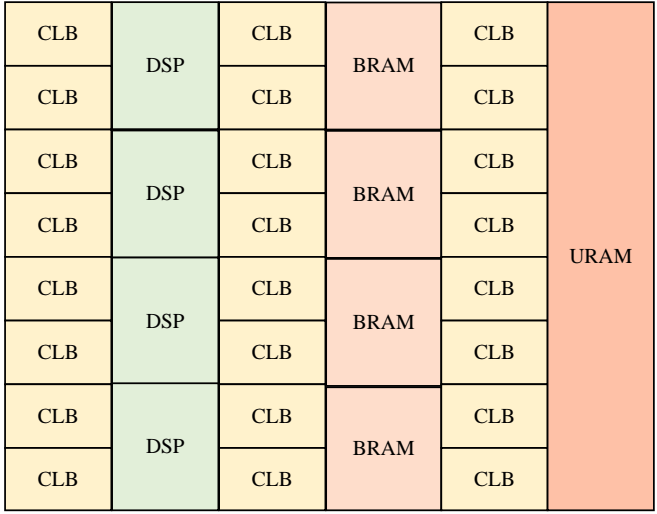

Fig. 1. A high-level view of a local area of the Xilinx FPGA layout.

## 3.2 Terminology and Notation

Hereafter, we define the terminology and notation used throughout this paper.

**FFCL Module** is defined as the netlist of a combinational circuit written in a hardware description language, such as Verilog.

**Compute kernel** is part of an application in the SDAccel framework and is associated with a given FFCL module running on the FPGA.

**Computational fabric** refers to a fabric that carries out the execution of the compute kernel. In this paper, we use computational fabric and FPGA interchangeably.

**Computational unit** refers to a block on the computational fabric that performs the execution of a logic operation such as AND, OR, XOR, and etc. In this paper, we use computational unit and DSP interchangeably.

**Compute Cycle** is the number of clock cycles which takes a compute kernel to execute a task.

**Logic Level** (depth) of a gate in a digital circuit is the maximum number of gates on any path from primary inputs of the circuit to the gate.

Some of the parameters used in this paper are summarized in Table 1.

## 4 PROPOSED METHOD

The overall flow of the proposed algorithm is as illustrated in Fig. 2. The input to the flow is a description of a FFCL module in Verilog format. Please note that the framework can be structured to accept any specification of an FFCL module as the input. ABC [1] and Yosys [21] and can be used to generate synthesizable Verilog code from any specification. NullaNet generates the FFCL module in Verilog format and it will be used as the upper stream engine as explained in Section 7. We first parse the Verilog netlist, synthesize the circuit using standard logic optimization techniques, primarily aimed at reducing the total gate count and depth of the circuit, and map the circuit to a customized cell library. The Boolean operations supported by the logic gates in the cell library, such as two input AND, OR, and XOR operation must be supported by the computational unit, i.e., DSPs on the FPGA. Next, the mapped circuit is levelized. Starting from the primary inputs of the module, each gate is assigned a logic level value that is one above the maximum logic level of its fanins. In other words, the logic level of each gate is determined as the maximum number of gates on any path from a primary input of the FFCL module to any of the inputs of the gate, in addition to one, accounting for the gate itself. As a gate with a specific logic level does not have any connections to any other gate with the same logic level, their operations can be executed simultaneously.

Table 1. Symbols used in this paper

| Term | Definition (bit-width values used in this paper) |
|---|---|
| $\lambda$ | Ratio of AXI data width to Address data width (36) |
| $\delta$ | Ratio of AXI data width to Input data width (10) |
| $\zeta$ | Ratio of AXI data width to Operation data width (85) |

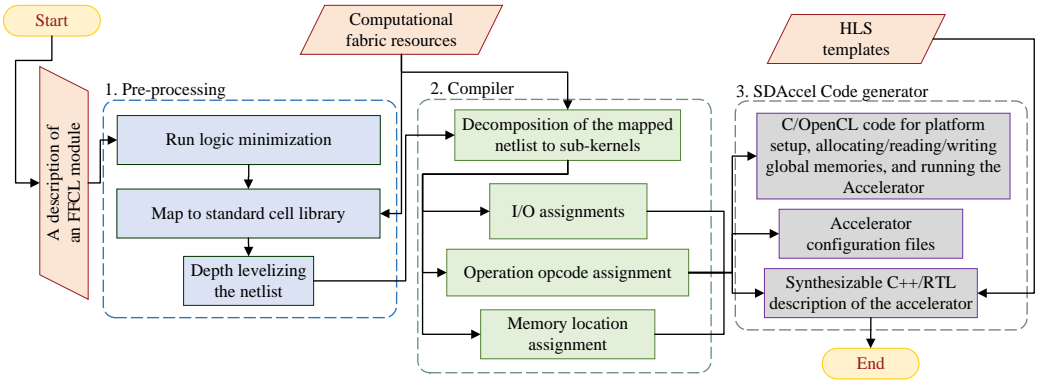

Fig. 2. An overview of the proposed framework.

Subsequently, the set of operations carried out by the gates in each logic level in the mapped netlist is decomposed into a set of sub-kernels; The number of sub-kernels is a function of the number of Boolean operations carried out at each level and the maximum number of available DSPs in the computational fabric. For instance, if the computational fabric is comprised of 1,000 DSPs, and in one level of an FFCL module there are 2,600 Booelan operations, the set of operations in this level is broken down into three sub-kernels. Note that since the FFCL module is levelized, in case there are enough DSPs available on a computational fabric, all the Boolean operations in different sub-kernels can be performed in parallel as there are no data dependencies between sub-kernels. The total number of sub-kernels required to implement the overall FFCL module using available DSPs in the computational fabric is calculated as summation of the number sub-kernels for each logic level, over all the logic levels. Consequently, the number of sub-kernels is a function of both the logic depth as well as the number of Boolean operations in each logic level of a FFCL module. Therefore, reducing the total number of Boolean operations and depth of the combinational logic during the logic synthesis and technology mapping steps are both paramount. Once the sub-kernels implementing a FFCL module are determined, we assign input and outputs of each sub-kernel specific numbers, representing the locations in memory where input operands are read from and outputs are saved to. These I/O assignments represent the locations in the memory of the computational fabric which can be implemented using available memory resources on FPGAs, such as BRAMs, URAMs, or look-up-tables (LUTs). Additionally, the operation corresponding to each DSP in each sub-kernel is configured. Finally, the assignment of memory locations and operation opcodes for each sub-kernel is saved in a JSON (javascript object notation) format, which will be later used to configure the operation of each DSP in the compute kernel in each time instance.

In the following sections, we first describe the proposed hardware accelerator and then present the compiler to support the efficient mapping on the proposed hardware accelerator.

## 5 HARDWARE ACCELERATION

HLS tools can significantly reduce development times through abstraction; however, they are seen as an additional step in the design flow, generating RTL code which must then go through the backend implementation flow, which is very time consuming. HLS design has a better abstraction for external modules and interfaces, making it easier and faster to implement complex control logic.

When designing systems on FPGAs, we wish to maximize the performance and efficiency of our circuits. This means making the best use of all types of resources available to us. Since designers generally write behavioral code that is then mapped by the implementation tools, performance and efficiency are controlled for the most part by these tools' capabilities. As architectures evolve with more complex resources, the tools have to work harder to make full use of them. While HLS tools allow higher-level design description, the final mapping remains the purview of the backend tools. If these cannot map general RTL code to exploit the architecture's capabilities, the resulting implementations can be inefficient. More importantly, the information contained in the high-level design description may help achieve this but be lost in the translation to generic RTL. For instance, current HLS tools cannot achieve as many and as much as fine-grained optimizations as RTL design tools, particularly for mapping logic function/gates to LUTs. Our experiments show that it takes several days for HLS tools to compile C++ code and generate the corresponding RTL code. Note that HLS tools are not able to map FFCL modules to DSPs. In other words, we can only explicitly force mapping mathematical operations (e.g., addition and multiplication) onto DSP blocks. Finally, in smaller designs where HLS tools can map the FFCL to LUTs, the generated design is about 20 times worse in terms of logic complexity compared to the specialized RTL design. Therefore, to fully realize advantages of both design approaches, We develop a mix of register-transfer level (RTL)/C++ descriptions of the Boolean logic expression accelerator where we use the RTL black-boxing feature of high-level synthesis (HLS) tools. This gives freedom to wrap RTL description of DSPs in optimized synthesizable C++ templates, acting as external modules or interfaces (like on-board double data rate memories, DDR), achieving low-latency accelerator designs on FPGAs.

The computation engine of the proposed accelerator is a custom fixed-function combinational logic fabric. This custom fixed-function generated in Section 6.1 differs for each Boolean expression. The reconfigurability of DSP blocks helps to reuse the same resources for implementing different Boolean expressions. Every DSP block can perform a 48-bit bitwise logic operation including AND, OR, NOT, NAND, NOR, XOR, and XNOR. This results in a SIMD scheme where we can perform the same operation using one OP-CODE for 48 different inputs. We store all input vectors in double data rate (DDR) memories, bring them all to on-chip ultra rams (URAM). Then, we divide the required computations in several rounds. In each round, a subset of the inputs (i.e., stored in the URAM) is transferred from URAM to its input block RAM (BRAM) for computation in each round. We also store opcode and addresses, where DSP registers must be read/written from/to BRAMs, in addition to input/output vectors. They are stored in opcode, address memory (cf. Addr. Mem. buffers in Fig. 3), and input vector buffers, respectively.

When we have all required data in the determined BRAM, we first read the data from designated locations for each computational unit, i.e., a DSP block, from the FPGA BRAMs and move them to the registers of the DSP blocks, which serve as the input for the custom combinational logic. The addresses for reading such data are accessible in the address memory buffer (cf. Fig. 3). The computation output is written to the output register files of the DSP block and then the BRAMs using fetched address. By iterating over the input vector BRAM and bringing different input data vectors to registers, performing the required Boolean operations, and storing the results from output registers to output BRAMs, the required computation for each Boolean expression is completed. This process is orchestrated by a control unit.

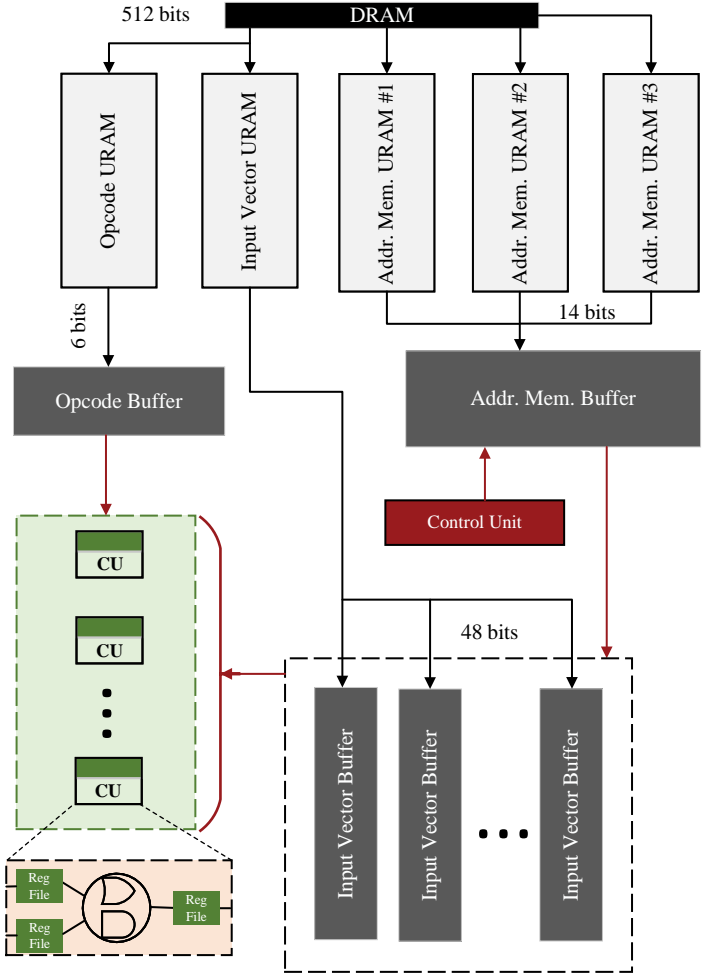

Fig. 3. The hardware architecture used in this paper. CU is the computational unit.

## 5.1 Memory Layout and Data Placement

To utilize the off-chip memory bandwidth, we group input data and addresses as well as opcodes before sending them to the on-chip memory. As shown in Fig. 3, the width of the packed data is 512 bits.

Required computations for each Boolean expression are done by iteratively reading and transferring data from the input BRAM to registers, executing custom combinational logic functions on the data, and storing results from output registers to BRAMs.

The width of the data stored in the registers and BRAMs for each Boolean expression is 48 to match the SIMD lines in the DSP block's logic unit. The width of opcodes are 6 bits while the width of addresses of data locations are 14 bits.

## 5.2 Hardware and Software Optimization

In the following sections, we will briefly discuss the optimizations considered in our hardware and software designs. To realize our design in FPGA, we use the Xilinx SDAccel and Vivado HLS, which provide a toolchain for programming and optimizing different applications on Xilinx FPGAs using a high-level language (C, C++ or OpenCL), as well as a runtime tool based on the OpenCL API, which is used by the host-side software to interact with the hardware accelerator known as hardware kernel.

*5.2.1 Burst Data Transfer.* We load all input data required for a FFCL's computations at once before any computations for the FFCL can begin. This lowers the FFCL's processing time, especially because now loading of input data, addresses, and opcodes can be done simultaneously (input data and addresses are stored in separate off-chip memory banks in the target FPGA board and are thus simultaneously accessible). Furthermore, we enable the full burst read/write of data from/to the memory banks and utilize the maximum possible burst size (512-bit width) and burst length (256 beats) allowable on the Advanced eXtensible Interface (AXI) bus of the target FPGA board. Specifically, to enable burst read of input data vectors, addresses, and opcodes, we allocate a *URAM* on the FPGA as shown in Fig. 3. Employing URAM blocks for implementing large buffers both prevents over-utilizing the BRAM blocks, and helps us achieve a more balanced utilization of resources in the target FPGA board.

*5.2.2 Double Buffering.* To further reduce the overhead of loading new FFCL data, during an FFCL's computations, we must preferably pre-fetch the data for the next FFCL so that the actual computations of the next FFCL can start earlier. To achieve this goal, we use the idea of double-buffering where we have two sets of buffers. One set of buffers contains all data (input data vectors, addresses, opcodes) required for the current FFCL's computations while the other is being filled with the data of the next FFCL that are read from the off-chip memory.

*5.2.3 Task pipelining.* Utilizing double-buffering offers an opportunity to pipeline tasks. We define two tasks, data movements and hardware kernel computations. In data movement, we transfer data from off-chip memory to URAMs and distribute addresses and opcodes to the pre-assigned BRAMs. In the kernel computation task, we iterate over vectors of the data and bring one vector of data from URAMs to BRAMs and perfom the on-chip operations. These tasks are explained in details in Section 6.2. Then, the control unit sends the necessary signal to read addresses from address memory buffers and loads the registers with the required data from pre-defined locations. Finally, control unit sends execution signals to computational units to perform required computations and stores back results to the BRAMs. The data transfer of the second task is shown with red arrows in Fig. 3.

*5.2.4 Multiple Parallel Accelerators.* Utilizing multiple parallel hardware kernels (i.e., accelerators) allows temporal parallelism, where the same hardware kernel processes different sets of data or different FFCL module. This is possible by enqueuing multiple hardware kernels in a pipelined manner. To achieve this, we follow the OpenCL programming paradigms. Enqueuing multiple accelerators happens through multiple OpenCL's clEnqueueTask commands in the host code, in a pipelined manner. clEnqueueTask is used to enqueue an accelerator to a command queue. OpenCL's clCreateCommandQueue API creates a command queue which keeps a track of queued tasks. In our design we use an out-of-order command queue to concurrently execute multiple hardware kernels.

## 6 COMPILER

As described in Section 4, our compiler parses, synthesizes, and levelizes the Verilog netlist to extract the set of operations carried out by the gates in each logic level in the mapped netlist. Then, our compiler decides on the total number of sub-kernels required to implement the overall FFCL module using available DSPs in a compute kernel considering the proposed hardware accelerator and available resources on the given FPGA. Finally, the assignment of memory locations and operation opcodes for each sub-kernel is determined, that will be later used to configure the operation of each DSP in the compute kernel in each time instance.

In the following subsections each of the above steps are detailed and examples are presented to further illustrate the proposed methodology.

### 6.1 Mapping to Logic Gates

In this section, we provide details about the process of mapping an FFCL module to a set of logic gates and the assignment of logic operations to DSP resources in the computational fabric, i.e, FPGA. In the first step, we use standard logic synthesis techniques to reduce the total gate count and the maximum logic depth of a circuit. The goal of this step is to transform a Verilog description of a FFCL module to a set of logic gates comprised of Boolean operations that can be carried out by the computational fabric, i.e, DSP devices on the FPGA. We use the ABC synthesis tool [1] to map a FFCL module to logic gates using the following commands:

```
1 $ resyn; resyn2; resyn2rs; compress2rs; st;
2 $ map; st; dch; map; st; dch; map
```

The first five commands reduce the size of the AND-Inverter Graph (AIG) representing the input logic network and are heuristic methods that optimize the AIG network. The *map* command is a *k*-feasible cut based mapper that maps the optimized AIG to a set of logic gates. The *st* command transforms the network back to the AIG form. The *dch* command performs AIG-based synthesis by repeatedly applying a sequence of technology-independent logic optimizations (using AIG rewriting rules).

Once the aforesaid sequence of commands is completed, the netlist is transformed to a set of 2-input logic gates, comprised of logic gates such as AND, OR, XOR, etc. Next, we assign logic levels to each gate. In this step, each logic gate is assigned a logic level $l$ as follows. The logic levels of the inputs of the FFCL module are set to 0. For each gate, the logic level is simply computed as follows:

$$l_i = 1 + \max_{j \in fanin_i} l_j \tag{1}$$

where $fanin_i$ represents the set of fanin gates for gate $i$. This step is straightforward and can be accomplished utilizing a breadth first graph traversal starting from primary inputs.

Because gates at the same logic level cannot have any connections to each other (i.e., from output of one to input of another), their operations can be carried out simultaneously. However, due to computational resource limitations i.e., the total number of available DSPs on the FPGA, it may be that not all the operations in a logic level can be executed in the same compute cycle. In such a case, the set of logic operations corresponding to a logic level are broken down into smaller subsets, called sub-kernels, and operations corresponding to different sub-kernels are carried out in a sequential manner.

We then compute the total number of sub-kernels at each logic level. If there are a total number of $n_i$ Boolean operations in logic level $i$ and there are a total number of $n_{DSP}$ DSPs in that target FPGA, the total number of sub-kernels of logic level $i$ will be $\lceil \frac{n_i}{n_{DSP}} \rceil$. Accordingly, the total number of sub-kernels required for mapping a FFCL module onto the FPGA is the sum (over all the logic levels) of the number of sub-kernels for each logic level. Hence, minimizing the total number of compute cycles is an optimization problem, involving both the total gate count and the max logic level of the circuit.

Next, Boolean operations corresponding to each sub-kernel are mapped to DSPs on the FPGA. Additionally, memory locations from which each DSP reads its inputs and locations where it writes its output must be determined. To do so, we first create a mapping between the inputs/outputs of each logic level with the BRAM locations on the FPGA. If there are a total number of $I$ inputs and $O$ outputs in a logic level, we require a total number of $I + O$ BRAM locations to store the data values corresponding to this logic level. The compiler first creates a simple mapping between the inputs/outputs of each logic level with BRAM locations on the FPGA. Next, for each sub-kernel, the assignment of Boolean operations to DSP devices as well as the locations for which data values are obtained from and written to are determined; Assuming there are $k$ Boolean operations in a sub-kernel, we require $k$ DSP devices; Since each DSP device reads two input values and generates one output value, we require $2 \times k$ memory locations for storing inputs of the sub-kernel and a total number of $k$ memory locations for storing output values. Initially, each Boolean operation is assigned to a DSP device, starting from 0 to $k-1$. Next, the memory locations where the inputs of the first DSP are obtained from are calculated using the aforementioned mapping between inputs of the logic level and the BRAM locations. Similarly, the BRAM location where the output of the Boolean operation should be written to is obtained using the same mapping. These memory addresses are then saved to BRAM locations; Consequently, during execution, the memory addresses where each DSP reads/writes data from/to is predetermined.

In the proposed methodology, the memory addresses of the inputs of the $p$th DSP are saved in BRAM locations $2 * p$ and $2 * p + 1$. Similarly, the address where the output of the first DSP is written is saved at BRAM location $2 * k$. The memory address where the output of the last DSP is written to is saved in BRAM location $3 * k - 1$. Finally, the look up table containing the assignment of BRAM locations is transferred to the FPGA and stored on BRAMs. At runtime, the host device reads the BRAM location assignments and transfers the contents of corresponding BRAM locations to the inputs registers of the DSP devices. Similarly, the outputs of DSP devices are stored to the pre-determined BRAM locations.

Consider an example where a 4-input AND gate is implemented using three 2 input AND gates; such a simple circuit can be broken down into two logic levels, where in the first logic level two parallel AND operations are carried out; Therefore, the first logic level has four inputs and two outputs; Initially, the inputs $(A, B, C, D)$ and outputs ($O1$ and $O2$) are simply mapped to BRAM locations 0 - 5. Next, we assume that the first AND operation (i.e., $O1 = A\&B$) is mapped to the DSP 1. Similarly, the second AND operation is mapped to DSP 2. Using the proposed methodology, the memory address of the inputs of DSP 1 are stored in BRAM locations $xFF00$ and $xFF01$. Similarly, the memory location where the output of DSP 1 should be written to is stored in BRAM location $xFF02$. According to the mapping between input/output signals of this logic level and BRAM locations, the contents of the BRAM locations $xFF00 - xFF02$ would be as follows: 0, 1, 4. Similarly, if the inputs/output of DSP 2 are mapped to BRAM locations $xFF03 - xFF06$, these memory locations contain the following values: 2, 3, 5. In this manner, during execution, each DSP simply reads the input values of a logic level, generates the corresponding output and writes the output to the pre-determined BRAM location.

An illustrating example is presented in section 6.3 to further elaborate the proposed methodology.

## 6.2 Computational Model

In this section, we formulate the number of compute cycles as a function of the topology of the neural network, i.e., its number of layers, number of filters per layer, number of Boolean operations per filter, numbers of inputs and outputs per filter, characteristics of the input dataset, and configurations of the computational fabric, i.e, the number of available resources such as DSPs, BRAMs, URAMs, LUTs, etc. for each compute kernel. Subsequently, we deploy the developed model and present a methodology to minimize the total number of compute cycles by optimizing the total number of DSPs on each compute kernel.

The number of compute cycles for carrying out computations of a FFCL is a function of multitude of parameters, including the topology of the neural network architecture, characteristics of the input dataset, and number and type of available resources in the computational fabric, i.e., the target FPGA. As described in Section 6.1, the process of mapping a FFCL to compute kernels involves two major tasks: (i) data movement and (ii) Boolean operations. The data movement task, which transfers data from the host memory to the device memory and vice versa, incurs a latency that is a function of the available resources on the computational fabric, such as sizes of the BRAMs and URAMs, and characteristics of the communication means of the computational fabric, such as the PCIe bandwidth and I/O data width. Before (and after) doing any compute kernel computations, the input (and output) data must also be transferred from the global computational fabric memory to the local compute kernel memory (and vice versa). For example, there are data transfers from/to BRAMs to/from DSP registers. The data movement task also accounts for this latency. The operation latency is associated only with that of processing the input vectors of the compute kernels and producing the output results.

One of the primary optimizations in the proposed flow is the parallel execution of the two aforesaid tasks. That is, one can transfer input vectors and do the assignment of memory addresses and opcodes for one compute kernel while simultaneously carrying out the Boolean operations on the input vectors for another compute kernel. Consequently, we can minimize the total compute cycles required for doing multiple compute kernels (which is the case in the NN inference) on a target computational fabric by pipelining these two tasks so that the data movement for one compute kernel overlaps in time with the Boolean operations of another compute kernel. With this pipelining scheme, the overall latency of executing $m$ FFCLs on a target computational fabric may be calculated as follows:

$$n_{cc,opt} = (m + 1) \times \max(n_{data\_moves}, n_{compute}) \qquad (2)$$

More precisely, the data movement cost consists of three parts: i) the cost of reading the input data and transferring operation codes that specify the Boolean operations to be performed on each DSP, ii) the cost of setting up addresses for loading from (and storing to) BRAMs before (and after) performing DSP computations, and iii) the cost of transferring the generated results. In our experiments, we observe that the latency associated with storing outputs is negligible. Furthermore, the cost of data movements associated with the transfer of memory addresses is larger than other data movement costs. So, we assign one (external) double data rate (DDR) SDRAM bank to the input vectors and operation codes and all other available DDR SDRAM banks to the memory addresses. Since we parallelize the two tasks of reading the input data and opcodes and the memory addresses load, the cost of data movements may be rewritten as:

$$vn_{data\_moves,opt} = \max(n_{read\_inputs\_opcode\_mem}, \\ n_{read\_addr\_mem}) \qquad (3)$$

Data movement consists of transferring input vectors, i.e., inputs of the FFCLs (which can be pixels of the input dataset or bit-packed vectors of the intermediate layers/filters), transferring Boolean operation codes that should be executed by the compute kernels, and transferring addresses of memory locations to load from or store values in the BRAMs which are being utilized by the compute kernels. The cost associated with bringing addresses of memory locations into the global on-chip memory and then distributing them to the local on-chip memories may be calculated as follows:

$$n_{read\_addr\_mem} = n_{AM\_DRAM\_to\_URAM} + (k-1) \times n_{AM\_URAM\_to\_BRAM} \quad (4)$$

where $n_{AM\_DRAM\_to\_URAM}$ is the latency of bringing addresses of memory locations from external DRAM into the URAM, $k$ denotes the total number of available DDR banks, which is 4 in our case (see Section 8 for more details regarding the target FPGA board), and $n_{AM\_URAM\_to\_BRAM}$ is the cost of transferring data from URAM into BRAM. Note that we use URAM (BRAM) as the global (local) on-chip memory in our design. $k-1$ is multiplying $n_{AM\_URAM\_to\_BRAM}$ since we have to distribute addresses of memory locations read from $k-1$ external DRAM in sequence to local on-chip memories.

Now then, we have:

$$n_{AM\_DRAM\_to\_URAM} = n_{subkernels} \times n_{subk\_addresses} \quad (5)$$

where $n_{subkernels}$ denotes the number of sub-kernels of a compute kernel and $n_{subk\_addresses}$ is the maximum number of addresses required for a sub-kernel.

Since there are two input and one output registers associated with each DSP, we need to bring three addresses for each DSP. In other words, $n_{subk\_addresses}$ is also the count of the registers holding input and output vectors for each computational unit (i.e., $3 \times n_{DSP}$). Because the bit width of the PCIe bus is much larger than number of bits assigned to each address, we can pack several such addresses into one bus transaction. Let $\lambda$ denote this packing number (cf. table 1). Moreover, we assign several memory banks to this task to reduce the total latency associated with task. Hence, the total latency associated with this task may be rewritten as:

$$n_{AM\_DRAM\_to\_URAM,opt} = \frac{n_{subkernels} \times n_{subk\_addresses}}{\lambda \times (k-1)}$$
$$= \alpha \times n_{subkernels} \times n_{DSP} \quad (6)$$

where

$$\alpha = \frac{3}{\lambda \times (k-1)} \quad (7)$$

The cost of transferring addresses from URAM to BRAM may be expressed as:

$$n_{AM\_URAM\_to\_BRAM,opt} = \frac{1}{2} \times n_{AM\_DRAM\_to\_URAM,opt} \quad (8)$$

Notice that $n_{AM\_URAM\_to\_BRAM,opt}$ is half of $n_{AM\_DRAM\_to\_URAM,opt}$ because we rely on true dual port BRAM with the ability to perform any combination of independent read or write operations in the same clock cycle. Aggregating (4) (6), and (8) yields:

$$n_{read\_addr\_mem,opt} = \frac{k+1}{2} \times n_{AM\_DRAM\_to\_URAM,opt}$$
$$= \beta \times n_{subkernels} \times n_{DSP} \quad (9)$$

where
$$\beta = \frac{k+1}{2} \times \alpha \qquad (10)$$

The cost associated with transferring the input vectors to each compute kernel plus that of transferring Boolean operation assignments to each DSP on a compute kernel can be estimated as follows:

$$n_{read\_inputs\_opcode\_mem,opt} = \lceil \frac{n_{input\_vectors} \times n_{fanin}}{\delta} \rceil$$
$$+ \lceil \frac{n_{subkernels} \times n_{DSP}}{\zeta} \rceil \qquad (11)$$

Where $n_{input\_vectors}$ is the total number of vectors that must apply to FFCL. $n_{fanin}$ is the number of primary fanins in the given FFCL. Notice that $\delta$ and $\zeta$ are added in equation (11) to capture the effect of data packing similar to equation (6). Finally, by combining equations (3), (9), and (11), the overall data movement cost can be expressed as:

$$n_{data\_moves,opt} = \max(\lceil \frac{1}{\delta} \times n_{input\_vectors} \times n_{fanin} \rceil +$$
$$\lceil \frac{1}{\zeta} \times n_{subkernels} \times n_{DSP} \rceil,$$
$$\beta \times n_{subkernels} \times n_{DSP}) \qquad (12)$$

The compute cycles associated with computing the output of each compute kernel for each input vector may be simplified as follows:

$$n_{compute\_one\_CK} = n_{loop\_subkernels} + n_{outputs} \qquad (13)$$

where $n_{outputs}$ is the cost associated with the transferring the generated outputs from local memory to the external (off-chip) memory. Recall that each FFCL is divided into a collection of subkernels. $n_{loop\_subkernels}$, which accounts for the cost of feeding the DSP registers with the proper data and then executing the logic operations in DSPs and storing back the results to the proper locations of local memory, is calculated as sollows:

$$n_{loop\_subkernels} = n_{subkernels} \times (n_{BRAM\_to\_DSP\_regs}$$
$$+ n_{exe\_logic\_ops}$$
$$+ n_{DSP\_reg\_to\_BRAM}) \qquad (14)$$

The latency of bringing the data to input registers of DSPs is as follow:

$$n_{BRAM\_to\_DSP\_regs} = \frac{2}{3} \times n_{subk\_addresses} = 2 \times n_{DSP} \qquad (15)$$

Note that 2/3 is because two out of three registers associated with a computational unit are accounted for input data and one is for the generated output data. We have observed that latency of this task dominates $n_{loop\_subkernels}$ when we increase the number of DSPs. Therefore, it is important to perform this task very efficiently. Since we already partition the memory for storing the addresses by a factor of $\lambda$, as seen in equation (8), we are able to parallelize the data transfer by a maximum factor of $\lambda$. However, we must access $\lambda$ data values at the same time. Unfortunately, the memory for storing the input vector is not and cannot be partitioned because there are no patterns in accessing the data which can dictate a reasonable partitioning solution. So, we use another trick and copy the input vector to multiple on-chip memories ($\lambda/2$ memories since we use dual-port memory) to increase the number of access lines. This results in a considerable reduction in the cost associated with feeding the new data to DSPs. Hence, after this optimization, equation (15) may be rewritten as:

$$n_{BRAM\_to\_DSP\_regs,opt} = \left\lceil \frac{2 \times n_{DSP}}{\lambda} \right\rceil \tag{16}$$

Moreover, the latency associated with copying an input vector to multiple on-chip locations must be added to equation (13):

$$n_{compute\_one\_CK} = n_{copy\_mem\_in} + n_{loop\_subkernels} + n_{outputs} \tag{17}$$

where

$$n_{copy\_mem\_in} = n_{fanin} \tag{18}$$

Similar to equation (16), the latency of transferring the results from DSP registers to the local on-chip memory is:

$$n_{DSP\_reg\_to\_BRAM,opt} = \left\lceil \frac{1}{2} \times n_{BRAM\_to\_DSP\_regs,opt} \right\rceil \tag{19}$$

Notice that $n_{DSP\_reg\_to\_BRAM,opt}$ is half of $n_{BRAM\_to\_DSP\_regs,opt}$ since there is only one output register assigned for each DSP.

Aggregating equations (14), (16), and (19), we can write:

$$n_{loop\_subkernels} = n_{subkernels} \times \left( \left\lceil \frac{2 \times n_{DSP}}{\lambda} \right\rceil + n_{exe\_logic\_ops} + \left\lceil \frac{1}{2} \times \left\lceil \frac{2 \times n_{DSP}}{\lambda} \right\rceil \right\rceil \right) \tag{20}$$

where $n_{exe\_logic\_ops}$ denotes the latency of a computational unit for executing a logic operation. If we have multiple data vectors, then the total latency of the compute part can be summarized as follows:

$$n_{compute} = n_{input\_vectors} \times n_{compute\_one\_CK} \tag{21}$$

$$= n_{input\_vectors} \times \left( n_{fanins} + n_{subkernels} \times \left( \left\lceil \frac{2 \times n_{DSP}}{\lambda} \right\rceil + n_{exe\_logic\_ops} + \left\lceil \frac{1}{2} \times \left\lceil \frac{2 \times n_{DSP}}{\lambda} \right\rceil \right\rceil \right) + n_{outputs} \right)$$

By aggregating the aforementioned equations in this section, one can obtain the final compute cycles for a single compute kernel carrying out the computation on an arbitrary number of input

vectors as follows:

$$\begin{aligned} n_{cc,opt} =& (m+1) \times \max\left( \max\left( \frac{1}{\delta} \times n_{input\_vectors} \times n_{fanin} \right.\right. \\ & \left. + \frac{1}{\zeta} \times n_{subkernels} \times n_{DSP}, \beta \times n_{subkernels} \times n_{DSP} \right), \\ & n_{input\_vectors} \times \left( n_{fanins} + n_{subkernels} \times \left( \lceil \frac{2 \times n_{DSP}}{\lambda} \rceil \right. \right. \\ & \left.\left.\left. + n_{exe\_logic\_ops} + \lceil \frac{1}{2} \times \lceil \frac{2 \times n_{DSP}}{\lambda} \rceil \rceil \right) + n_{outputs} \right) \right) \end{aligned} \quad (22)$$

where $m$ denotes the number of FFCLs as stated above.

Additionally, the number of subkernels associated with each FFCL is a function of the number of logic levels, the number of Boolean operations per logic level, and the number of available DSPs as follows:

$$n_{subkernels} = \sum_{l=1}^{L} \left\lceil \frac{n_{gates}^{l}}{n_{DSP}} \right\rceil \quad (23)$$

where $l = 1 \ldots L$ denotes the logic levels for each FFCL and $n_{gates}^{l}$ is the number of gates in logic level $l$. As seen from (22), the number of compute cycles is a non-linear function of the number of availablecomputational units on the computational fabric.

## 6.3 Illustrating Examples

In this section, we outline the process of mapping two FFCL modules to a set of computational units (DSPs) on a computational fabric (FPGA), using the proposed methodology.

Consider two FFCL modules, each implementing a Boolean expression of 4 input values, as depicted in Figures 4 and 5. In the proposed flow, initially, each of the designs is levelized, i.e., operations in the design are assigned to specific logic level based on as-soon-as-possible scheduling strategy. As a result, each Boolean operation is assigned a logic level $l$, where $l$ denotes the smallest value larger than the logic level of all the input operands. Consequently, designs 1 and 2 each comprise 2 and 3 logic levels. Next, according to the number of available resources in the computational fabric (i.e., the number of available computational units), the set of Boolean operations in each logic level are clustered into one or multiple subkernels. In this example, there are only two computational units available. Consequently, each level in design 1 is divided to one subkernel, whereas level 1 in design 2 is divided into two subkernels. Due to data dependencies, the operations of all subkernels of a level should be completed before subkernels of the next level can be launched. The number of subkernels for all the levels of an FFCL module determines the number of clock cycles it takes for the computational fabric to compute the output result of the FFCL for one vector of its input values. Accordingly, the computations for designs 1 and 2 are completed within 2 and 4 cycles, respectively.

Now consider each of the designs $g_1$ and $g_2$. The contents of the input data buffer, opcode buffer, and address memory buffer for realizing function $g_1$ are depicted in Table 2. In our proposed methodology, indices 0 and 1 of the input data vector are always filled with constant values of 0 and 1, representing constant values in Boolean expressions. Furthermore, the compiler populates the next four indices with the values of inputs of the FFCL module, i.e., values of inputs $a - d$. As outlined before, the compiler creates a mapping between internal values, i.e., nodes in the graph

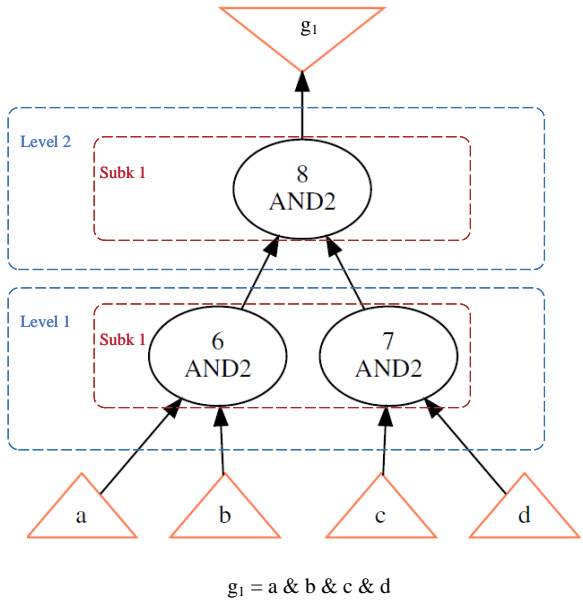

Fig. 4. Illustration example 1.

Table 2. The contents of the input data buffer, opcode buffer, and address memory buffer for realizing function $g_1$ (c.f. fig. 4).

| Index | Data Vec. Buf. | Addr. Mem. Buf. | Opcode Buf. |
|---|---|---|---|
| **0** | 0x0000 | [ 2, 3, 4, 5 ] | [AND,AND] |
| **1** | 0xFFFF | [ 6, 7, 0, 0 ] | [AND, NOP] |
| **2** | a | [ 6, 7, 0, 0 ] | -* |
| **3** | b | [ 8, 0, 0, 0 ] | - |
| **4** | c | - | - |
| **5** | d | - | - |
| **6** | w1 = a & b | - | - |
| **7** | w2 = c & d | - | - |
| **8** | out = w1 & w2 | - | - |

*Values represented with "-" means that there is no data in corresponding indices. It is obvious that address memory and opcode buffers are smaller than data vector buffers in this example.

representing the Boolean function, and locations in the memory. Consequently, the intermediate values corresponding to nodes $w_1$ and $w_2$ are mapped to locations 6 and 7 of the input data vector buffer. Finally, the output value is stored in location 8 of the data vector buffer. As observed, the total size of the data vector buffer for realizing a FFCL is calculated as the total number of nodes of the directed acyclic graph (DAG) representing the Boolean expression.

The contents of the address memory buffer are also listed in Table 2. As shown, the cardinality of each vector in the address memory buffer is equal to 2× the number of computational units, as each

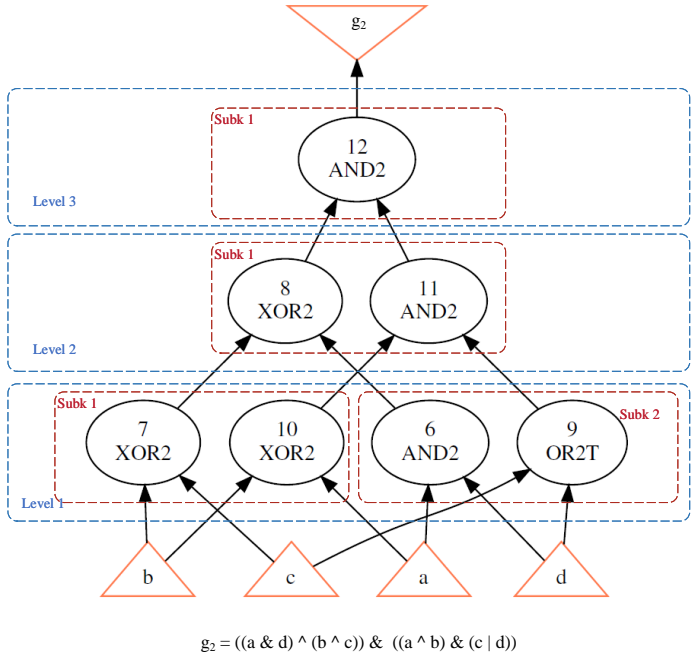

$g_2 = ((a \& d) \wedge (b \wedge c)) \& ((a \wedge b) \& (c | d))$

Fig. 5. Illustration example 2.

Table 3. The contents of the input data buffer, opcode buffer, and address memory buffer for realizing function $g_2$ (c.f. fig. 5).

| Index | Data Vec. Buf. | Addr. Mem. Buf. | Opcode Buf. |
|---|---|---|---|
| **0** | 0x0000 | [ 3, 4, 3, 2 ] | [XOR_OP, XOR_OP] |
| **1** | 0xFFFF | [ 6 (w1), 7 (w2), 0, 0 ] | [XOR_OP, OR_OP] |
| **2** | a | [ 5, 2, 5, 4 ] | [XOR_OP, AND_OP] |
| **3** | b | [ 8 (w3), 9 (w4), 0, 0 ] | [AND_OP, NOP] |
| **4** | c | [6, 8, 7, 9] | - |
| **5** | d | [ 10 (w5), 11 (w6), 0, 0 ] | - |
| **6** | w1= b ^ c | [ 10 , 11, 0, 0 ] | - |
| **7** | w2 = b ^ a | [ 11 (out), 0, 0, 0] | - |
| **8** | w3 = d ^ a | - | - |
| **9** | w4 = d \|\| c | - | - |
| **10** | w5 = w1 ^ w3 | - | - |
| **11** | w6 = w2 & w4 | - | - |
| **12** | out = w6 & w5 | - | - |

computational unit requires two operands. In the first compute cycle, the first computational unit reads the value of the operands from indices 2 and 3 of the memory while the second computational unit obtains its operands by reading input data vector buffer locations 4 and 5, respectively (cf.

Table 2, Column address memory buffer, Index 0). Subsequently, computational units write the output values into locations 6 and 7 of the data vector buffer.

In the second compute cycle, the first computational unit performs operation on values obtained from data vector buffer locations 6 and 7 and stores the output in data vector buffer location 8. In this cycle, the second computational unit does not perform any operations. As shown in Table 2, the Boolean operation carried out by each computational unit is stored in the opcode memory, in a similar fashion to address memory buffer, albeit with the difference that the cardinality of the vectors in opcode memory is equal to number of of computational units.

Similar to design $g_1$, the contents of the memories for design $g_2$ is listed in Table 3. As shown, since the number of subkernels and operations carried out by FFCL module $g_2$ is larger than $g_1$, the size of the memories are increased.

## 7 APPLICATION OF PROPOSED METHOD TO NN INFERENCE

In this section, we explore the application of the proposed method to FFCL-mapped neural networks are discussed. As an example, we first describe Nullanet as natural fit for our framework and then present the compiler optimizations specific to neural network inference. We choose Nullanet as the method for generating FFCL modules since its newer verison, NullaNet tiny [7] outperforms LogicNets. LUTNet is specialized for mapping to LUTs and is extended from general XNOR operation while our goal is to to map a large Boolean functions with many input variables and product terms. We find such Boolean functions in implementing neural network using NullaNet method.

### 7.1 NullaNet

A summary of the NullaNet [8] flow is as follows. NullaNet first discretizes input and output activations of artificial neurons to binary values while training a DNN. Next, it forms Boolean specifications for the said neurons either by enumerating all their possible input combinations and calculating their corresponding outputs (i.e., realization based on input enumeration) or through applying all training data points to the neural network and for each neuron, recording values of the binary inputs and outputs encountered when processing each data point (i.e., realization based on incompletely specified functions (ISFs)). Realization based on input enumeration implements the exact same function as the one realized using MAC operations. However, it is only applicable to neurons with a small number of inputs (e.g. 14 inputs or less).

Realization based on ISFs, on the other hand, samples the algebraic function that represents each neuron and transforms that algebraic function to a Boolean function that approximates it. This approach is suitable for implementing neurons designed for state-of-the-art neural networks which include tens to hundreds of inputs. In such neurons, the input space is huge and the samples only represent a tiny fraction of the input space that matters to the DNN, hence the approximation. After finding the Boolean specification of each neuron, NullaNet employs logic synthesis to find a near-optimal realization of each neuron by optimizing its corresponding Boolean function. During inference, the output of each neuron, which is normally calculated by applying dot product, batch normalization, and the application of an activation function, is simply calculated by a number of logic operations that were found during the logic synthesis step. This paradigm shift not only enables significant savings in computations, it also eliminates the need to access a neuron's parameters during inference, which leads to substantial savings in energy and latency.

### 7.2 Optimization Problem

When running a given dataset (e.g. CIFAR-10) through a set of FFCLs and carrying out computations associated with a given neural network (e.g. VGG-16), minimizing the total compute cycles (determined by the summation of compute cycles for each neural network layer) is of paramount importance. Assuming the computation of each layer is carried out in a sequential manner (i.e., we finish computations of one layer before starting computations of the next layer), the total compute cycles for computing the outputs of a set of FFCLs can be formulated as follows:

$$n_{cc,opt}^{nn} = \sum_{i=1}^{M} n_{filter}^{i} \times n_{cc,opt}^{i} \quad (24)$$

In the above example, $i = 1 \ldots M$ represents various layers of a neural network, $n_{filter}^{i}$ denotes the number of filters in each layer $i$ of the network, and $n_{cc,opt}^{i}$ is the maximum compute cycles for applying a filter to input feature map of each layer $i$ of the network.

As previously mentioned, depending on the available resources on the computational fabric, one can reduce the total cycle count by launching multiple compute kernels in parallel, all executing the same Boolean function on the set of input vectors. In this manner, the total number of compute cycles for a set of compute kernels implementing a FFCL module is greatly reduced by the number of parallel executing compute kernels (denoted by $n_{parallel\_factor}$) as follows:

$$n_{cc,opt}^{tot} = \frac{n_{cc,opt}^{nn}}{n_{parallel\_factor}} \quad (25)$$

Because the number of input vectors, the number of outputs, and fanin count corresponding to each filter in each layer of the network is determined by the characteristics of the input dataset, network topology, and deployed pruning algorithms, minimizing the total compute cycles is a multi-tier optimization. However, assuming a fixed network, dataset, and compression algorithm, one can obtain the optimal number of computational units, i.e., DSPs, by minimizing equation (24):

$$\min \sum_{i=1}^{M} \lceil \frac{n_{filter}^{i} \times n_{cc,opt}^{i}}{n_{parallel\_factor}} \rceil$$
$$\text{subject to } n_{DSP} \leq N^{DSP} \quad (26)$$

where $N^{DSP}$ denotes the number of DSPs in the target FPGA.

Using the above problem formulation, the design space for the given neural network (associated with different values of computational units) is explored by using a simple binary search algorithm in order to find the best solution for the given neural network (see Fig. 6).

## 8 SIMULATION RESULTS

For evaluation purposes, we targeted a high-end Virtex® UltraScale+ FPGA (Xilinx VU9P FPGA [*], which is available in the cloud as the AWS EC2 F1 instance). This FPGA platform includes 64 GiB DDR4 ECC protected memory, with a dedicated PCIe x16 connection. There are four DDR banks. This FPGA contains approximately 2.5 million logic elements and approximately 6,800 DSP units[†]. We use SDAccel 2019.1 for mapping to the FPGA and set the target frequency to 250 Mhz.

First, we assess the proposed model described in the section 6.2. Then, we discuss the efficiency of proposed accelerator presented in the section 5. Finally, we evaluate our proposed method on two well-known CNN, i.e., VGG16 [15] and LENET-5 [5] and two commonly used computer-vision datasets for object recognition i.e., the CIFAR-10 [4] and MNIST [3] datasets.

---

[*]The full part name is xcvu9p-flgb2104-2-i.
[†]https://aws.amazon.com/education/F1-instances-for-educators/

## 8.1 Efficacy of Model Used in Compiler Optimization

To show the efficacy of the proposed analytical model used in the compiler optimizations, we compare the expected latency results from our compiler to actual results after running FPGA. For our comparison, we employ the parameterized analytical modeling tool presented in section 6.2 and extract the expected latency. Then, we evaluate the affinity of our model to actual realization according to latency values vs usage of different number of DSPs. Fig. 6 shows a comparison between our proposed model used in the compiler and actual hardware implementation in terms of the achieved performance for layer 7 of VGG16 network. Our model can predict actual performance by less than 10% error. This is mainly because we assumed the worst case scenario in our calculations i.e., that all DSPs are used when doing all sub-kernels of all FFCLs, which may not be the case. This pessimistic assumption affected both data movements and computations tasks calculations.

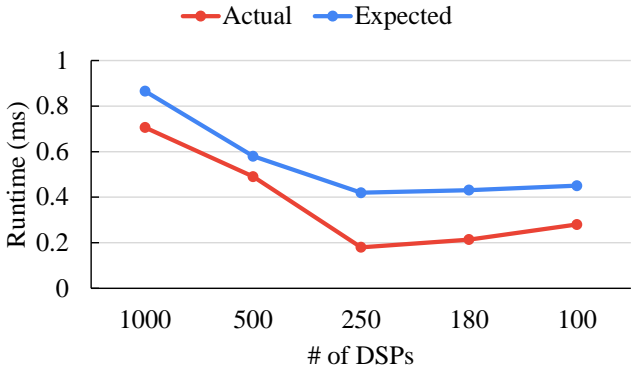

Fig. 6. A comparison between our proposed model used in the compiler and actual hardware implementation in terms of the achieved performance for layer 7 of VGG16 network.

As can be seen in Fig. 6, the design space can be inspected through a binary search algorithm to find the best solution in terms of the performance. The best solution is the one that minimize the latency represented by the number of cycles needed to compute all FFCLs by varying the number of computational units required for computations. Reducing the number of computational units boosted the performance since the data movements cost decreases. We observed that the total latency increase after a certain point by reducing the number of computational units because the computation cost became dominant.

## 8.2 Analytical Comparison: Memory Communications vs Computations

We analyze and quantify the strengths of our proposed accelerator in terms of latency (e.g., number of spent cycles) of memory communication vs computations. Fig. 7 depicts the proportional percentage of latency spent on memory communications and computations phases. Our proposed compilation and mapping algorithms achieve a balance between memory communication and main computation latency in various number of utilized DSPs. Computation latency outweights the memory communication latency when the number of DSPs are reduced. This is expected since we reduce the total number of resources, so the computations takes longer to finish.

Fig. 8 shows the importance of achieving a balance between data movements and computation latency in our proposed framework. If we do not consider this balance between both tasks in our design, we have to accept sub-optimal solutions because we pipeline the tasks as explained in Section 5.2.3. In other words, since the throughput of a pipeline cannot be better than that of its slowest stage (i.e., a task latency in our case), the designer and the compiler should try to divide the work and resources among the stages so that the tasks take the same to time to be completed. Please note that we employed a *single computation engine architectures*. As its name implies, this approach utilizes a generic accelerator architecture comprising a single computation engine that is used for the computation of all NN layers.

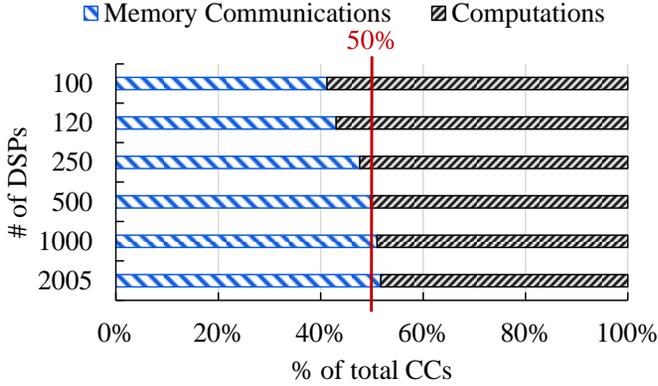

Fig. 7. Distribution of latency spent for memory communications and computations phases. The 50% threshold is marked to improve readability.

### 8.3 Comparison Between MAC-based, XNOR-based and Nullanet-based Implementation of CNNs

We use VGG-16 with CIFAR-10 dataset as a case study. The 16 in VGG16 refers to its 16 layers that have weights. This network is a huge network and it has about 138 million parameters. We implement intermediate convolutional layers 2-13 in VGG-16 using the proposed framework and fixed-function combinational logic functions.

As a baseline for the state-of-the-art generic MAC array-based accelerator for the layers realized using conventional MAC calculations, we used the open-source implementation of [16] with some improvements including transferring all weights required for the computation of the layer from the external memory into BRAMs, where these weights get reused for calculations corresponding to different patches of input feature maps. Furthermore, partial sums of accumulation for processing the output of a filter/neuron are also stored in the register file of the same processing element. Considering these improvements, we reduce the latency of VGG-16 inference employing the generic MAC array-based accelerator. We use FINN [19] for our XNOR-based baseline and replace the LUT-based XNOR unit with a DSP-based XNOR unit in the Matrix-Vector–Threshold Unit (MVTU) in order to have a fair comparison. In other words, all implementations used in this paper utilize the same set of hardware resources.

Resource utilization for all implementations for different number of DSP blocks on CIFAR-10 is reported in Table 4. Note that resource utilization for different implementations are the same because the first and the last layers have full-precision and thus they require an array of MAC units for their computations. Fig. 9 shows the achieved performance of all three types of implementations in different number of DSPs. We named our proposed accelerator as *NullaDSP*. As illustrated in the figure, the XNOR implementation achieves significant saving since we keep all intermediate and weights on on-chip memories and there is no cost associate with off-chip memories. Whereas this is not the case for MAC-based and our proposed implementations and memory communications constitutes a great proportion of the total latency. The latency of the MAC- and XNOR-based implementations is increased when the number of DSPs are reduced while the latency of our implementation forms a Pareto shape (cf. Fig. 6). The proposed implementation can achieve better performance (runtime of 2.99 ms) with less number of DSPs compared to MAC-based implementation with 1024 DSPs (5.72 ms). Using the proposed method, the total latency for VGG16 on CIFAR-10 datasset is reduced by around 2x compared to employing the MAC array accelerator design. Furthermore, the obtained accuracies using both of these approaches are relatively close. The model accuracy when all layers are mapped using MAC array accelerator design is obtained as 93.04%, while it is obtained as 92.26% when layers 2-13 are mapped using Nullanet method. It is 89.61% in the case of XNOR-based implementation. The compute-to-communication ratio for MAC array accelerator increases by assigning more DSPs to the compute while the trend is for NullaDSP accelerator. The best achievable latency utilizing the entire device is 0.14ms using 4,127 DSPs running at a clock frequency of 250 MHz.

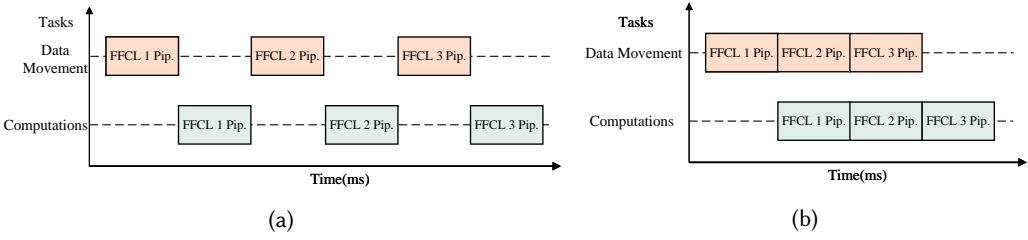

Fig. 8. The overall data flow of the proposed implementation for NN inference (a) without pipelining between tasks within a layer, (b) with task pipelining.

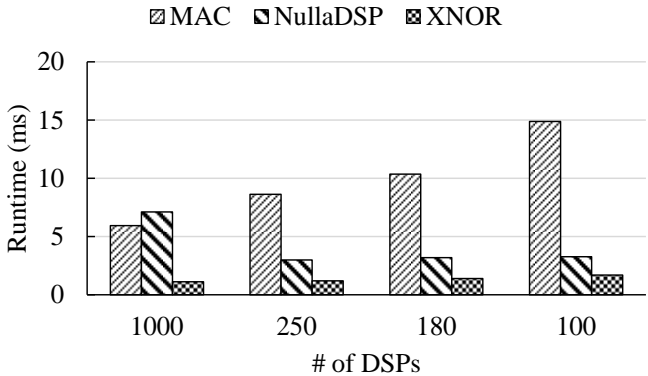

Fig. 9. Comparison between MAC-, XNOR-, and NULLANET-based implementation (e.g., NullaDSP). The main resource for the computations in all three type of the implementations are DSP blocks.

Table 4. Resource utilization for all implementations for different number of DSP blocks on CIFAR-10.

|  | Large[†] | Medium[†] | Small[†] | Tiny[†] |
|---|---|---|---|---|
| DSP | 1000 | 250 | 180 | 100 |
| FF(%) | 30K (1%) | 30K (1%) | 30K (1%) | 30K (1%) |
| LUT(%) | 77K (6%) | 75K (6%) | 74K (6%) | 74K (6%) |
| BRAM(%) | 3162 (73%) | 1848 (42%) | 1686 (39%) | 1512 (35%) |
| URAM(%) | 55 (5%) | 55 (5%) | 55 (5%) | 55 (5%) |
| Target Frequency(Max) (MHz) | 250 (253) | 250 (282) | 250 (291) | 250 (299) |

[†]We named implementations with 1000, 250, 180, 100 DSPs as Large, Medium, Small, and Tiny designs, respectively.

Furthermore, Fig. 10 shows the performance results of our flow for LENET-5 networks on MNIST, alongside a comparison with the other two type of implementations. The network has 5 layers with learnable parameters and hence named LENET-5. It has three convolution layers with two max pooling units in between. After the convolution and max pooling layers, we have two fully connected layers. It was successfully applied for identifying handwritten numbers provided in MNIST dataset.

As shown in the figure, our proposed implementation achieved a better performance for LENET-5 in terms of obtained inference latency compared to other types of implementations (e.g., achieving up to 20% latency improvement for 140 DSPs), while using comparable amount of resources on the target FPGA. The reason that our implementation outperforms the XNOR-based implementation in the case of LENET-5 is that parallelization potentials for XNOR-based implementation is not that high. To explain the reason in detail, first we describe the connection between dataflow and spatial loop unrolling in neural networks compilers.

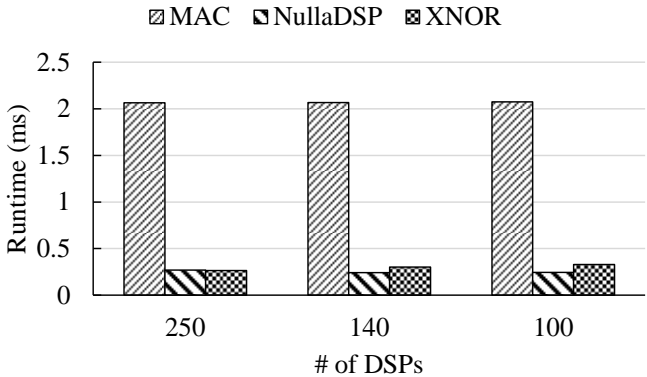

Fig. 10. Comparison between MAC-, XNOR-, and NULLANET-based implementation. The main resource for the computations in all three type of the implementations are DSP blocks.

The computational flow for a convolutional layer in CNN can be represented by a six-level nested loop (seven-level nested loops when considering the iteration over images in a mini-batch) known as a computational block [17]. We describe the dataflow of an accelerator through the mapping of particular loops to the parallel computation structures. In other words, the data communication pattern is determined by which loops are spatially unrolled in hardware, and which are not. In XNOR-based implementation, FINN [19] unrolls the input channel and output channel loops (i.e., weight stationary pattern), where the weights stay and are reused within the same PEs, but the inputs and outputs are spatially broadcast or accumulated. Since the number of input/output channels are limited in the LENET-5 architecture, then unrolling them and mapping to parallel computation structures do not improve the performance.

## 9 CONCLUSION

In this paper, we presented a novel design and optimization methodology for the compilation and mapping of fixed function neural networks to digital signal processors (DSPs) on the FPGAs employing high-level synthesis flow. The proposed methodology maps the fixed function combinational logic blocks to a set of Boolean functions and Boolean operations are mapped on DSP devices rather than look-up tables (LUTs) on the FPGAs to take advantage of the high performance, low latency, and parallelism of DSP blocks.